\documentclass[twocolumn, superscriptaddress, aps, prb]{revtex4-2}
\newif\ifGALLEYversion\GALLEYversiontrue
\usepackage{ifthen}
\usepackage{graphicx}
\usepackage{amsmath}
\usepackage{xcolor}

\graphicspath{{figures/}}

\ifthenelse{\boolean{GALLEYversion}}{
    \setlength{\marginparwidth}{2.7in}
    \setlength{\marginparsep}{0.5in}
    \def\ttm#1{\marginpar{\small TT: #1}}
}{
    \def\ttm#1{\relax}
}

\begin{document}
%%%%%%%%%%%%%%%%%%%%%%%%%%%%%%%%%%%%%%%%%%%%%%%%%%%%%%%%%%%%%%%%%%%%%%%%

%%%%%%%%%%%%%%%%%%%%%%%%%%%%%%%%%%%%%%%%%%%%%%%%%%%%%%%%%%%%%%%%%%%%%%%%
\title{Van der Waals Density Functional for Molecular Crystals}

\author{Trevor Jenkins}
\affiliation{Department of Physics, Wake Forest University, Winston-Salem, NC 27109, USA.}
\affiliation{Center for Functional Materials, Wake Forest University, Winston-Salem, NC 27109, USA.}

\author{Kristian Berland}
\email[E-mail: ]{kristian.berland@nmbu.no}
\affiliation{Department of Mechanical Engineering and Technology Management, Norwegian University of Life and Sciences, Norway}

\author{Timo Thonhauser}
\email[E-mail: ]{thonhauser@wfu.edu}
\affiliation{Department of Physics, Wake Forest University, Winston-Salem, NC 27109, USA.}
\affiliation{Center for Functional Materials, Wake Forest University, Winston-Salem, NC 27109, USA.}

\date{\today}

\begin{abstract}
Since the development of the nonlocal correlation functional vdW-DF, the
family of van der Waals density functionals has grown to better describe
a wide variety of systems. A recent generation of the vdW-DF family,
vdW-DF3, featured a newly-constructed form of the nonlocal correlation
that more accurately modeled molecular dimers, layered structures, and
surface adsorption.  However, it also revealed an intrinsic tradeoff in
vdW-DF3's parametrization and inflexibility of exchange in the
generalized gradient approximation (GGA), limiting its accuracy for
molecular crystals. In this paper we propose a new optimization of
vdW-DF3 that is tailored to 3D molecular crystals. This functional,
called vdW-DF3-mc, contains a new, tunable form of the exchange
enhancement factor with parameters that directly correspond to
physically relevant qualities. In addition, within the nonlocal
correlation, we prioritize smoothness of the kernel switching function
as a means of restoring flexibility to vdW-DF3's design. Testing
vdW-DF3-mc on several benchmark sets, we achieve highly accurate
energetics and geometries for molecular crystals.  This is particularly
evident for the case of polymorphs of ice, for which errors in the
volume and cohesive energy are on the order of only 1\%, indicating very
promising performance for important subcategories of molecular crystals,
such as polymorphism and hydrogen-bonded solids.
\end{abstract}

\pacs{71.15.Mb, 31.15.ae, 33.15.Dj}
% 71.15.Mb    DFT, LDA, gradient and other corrections
% 31.15.ae    Electronic structure and bonding characteristics
% 33.15.Dj    Interatomic distances and angles 

\maketitle
%%%%%%%%%%%%%%%%%%%%%%%%%%%%%%%%%%%%%%%%%%%%%%%%%%%%%%%%%%%%%%%%%%%%%%%%

%%%%%%%%%%%%%%%%%%%%%%%%%%%%%%%%%%%%%%%%%%%%%%%%%%%%%%%%%%%%%%%%%%%%%%%
\section{Introduction}\label{vdw-df3-mc:introduction}
%%%%%%%%%%%%%%%%%%%%%%%%%%%%%%%%%%%%%%%%%%%%%%%%%%%%%%%%%%%%%%%%%%%%%%%

The ubiquity of van der Waals interactions in nature is well documented,
and an accurate quantum mechanical description of these interactions is
therefore vital for reliable modeling of a wide variety of systems.  But
traditional density functional theory (DFT) has struggled to incorporate
these interactions, though much research has been dedicated to amending
that fact \cite{Dion_2004:van_waals, Klimes_2012:perspective_advances,
Thonhauser_2015:spin_signature, Langreth_2009:density_functional,
Berland_2015:van_waals, Vydrov_2009:nonlocal_van,
Vydrov_2010:dispersion_interactions,
Vydrov_2010:nonlocal_van,Grimme_2004:accurate_description,
Grimme_2007:density_functional, Grimme_2011:density_functional,
Grimme_2016:dispersion_corrected, Tkatchenko_2009:accurate_molecular,
Tkatchenko_2012:accurate_efficient,
Ambrosetti_2014:long_range_correlation, Ambrosetti_2016:wavelikeMBD,
Szalewicz_2012:symmetry-adapted_perturbation,
Burns_2011:density-functional_approaches}. The family of van der Waals
density functionals, beginning with the work of Dion \emph{et al.}\ in
2004 \cite{Dion_2004:van_waals}, represents one of the key developments
in this field. Through the use of a fully nonlocal correlation
functional which obeys exact physical constraints, vdW-DF1, for the
first time, provided a density-based description of dispersion
interactions for general geometries. Since then, the family of van der
Waals density functionals (vdW-DF) has grown significantly, including
the development of a third generation in 2020, vdW-DF3
\cite{Chakraborty_2020:next-generation_nonlocal}.  Building on the
earlier conceptual vdW-DF-C6 functional \cite{berland_2019:van_waals},
vdW-DF3 allows for re-parametrization of the plasmon dispersion model,
which determines the strength of dispersion forces.  Two optimizations
of vdW-DF3 achieved a high degree of accuracy for many different
dispersion-dominated systems. But during the development of vdW-DF3, it
was found that there are competing interests between different system
types, which limited its all-around performance. In particular, this
limited vdW-DF3's accuracy with respect to 3D molecular crystals. More
recently, we have traced the cause behind this competition: i.e.,
different classes of systems and interactions possess a very
characteristic profile of their electron density gradient
\cite{Jenkins_2021:reduced-gradient_analysis, Jenkins2024_Reduced}.
Thus, a functional can provide accurate predictions for two distinct
classes at the same time as long as their profiles are not significantly
overlapping. But, when having two classes with competing profiles, any
optimization will have to favor one class over the other.

To fill this important niche, we put forth a new pragmatic optimization
within the vdW-DF3 framework tailored to molecular crystals, which we
name vdW-DF3-mc. To leverage the insight provided by our
reduced-gradient analysis \cite{Jenkins_2021:reduced-gradient_analysis,
Jenkins2024_Reduced}, we have designed a new form of the enhancement
factor for the exchange in the generalized gradient approximation (GGA),
$F_{\rm x}(s)$. This form allows us to directly target qualitative
properties of individual or groups of systems, and in tandem
re-parameterize the vdW-DF3 correlation.  The reference sets for
re-parameterization were chosen as the X23 set of molecular crystals,
alongside 2 layered systems, and 24 molecular dimers at different
separations. Layered systems were included to represent molecular solids
with interplanar dispersion interactions such as $\pi$--$\pi$ stacking,
while the molecular dimers were included to represent porous 3D
materials with long-range interactions, such as the widely-studied
covalent organic frameworks (COFs) and hydrogen-bonded organic
frameworks (HOFs) \cite{liu_2010:stuctural_design,
li_2020:hydrogen_bonded, yu_2023:hydrogen_bonded,
geng_2020:covalent_organic, ge_2024:comprehensive_review}.  We find that
vdW-DF3-mc yields greatly improved accuracy over its predecessors for a
wide variety of 3D solids, both within and beyond our optimization set.
For energetics and geometries of conventional molecular crystals, its
accuracy is comparable to---and in some cases surpasses---that of the
force-field corrected PBE-D3 \cite{Grimme_2010:consistent_accurate},
which has previously demonstrated very good performance for the X23
\cite{moellmann_2014:dft-d3_study}.  Finally, while not specifically
optimized for ice, vdW-DF3-mc provides strikingly accurate results for
both the energy and structure of ice polymorphs.

%%%%%%%%%%%%%%%%%%%%%%%%%%%%%%%%%%%%%%%%%%%%%%%%%%%%%%%%%%%%%%%%%%%%%%%
\section{Theory}\label{vdw-df3-mc:theory}
%%%%%%%%%%%%%%%%%%%%%%%%%%%%%%%%%%%%%%%%%%%%%%%%%%%%%%%%%%%%%%%%%%%%%%%

%%%%%%%%%%%%%%%%%%%%%%%%%%%%%%%%%%%%%%%%%%%%%%%%%%%%%%%%%%%%%%%%%%%%%%%
\subsection{Constructing Exchange in the GGA}

In van der Waals density functionals, the exchange correlation can be
expressed as a sum of the GGA exchange, the local density
approximation of correlation, and the nonlocal correlation. That
is,
\begin{equation} \label{eq:Exc}
E_{\rm xc}[n] = E_{\rm x}^{\rm GGA}[n] +
E_{\rm c}^{\rm LDA}[n] + E_{\rm c}^{\rm nl}[n]\,.
\end{equation}
Here, the GGA exchange and nonlocal correlation are particularly
important for an accurate description of binding in van der Waals
complexes \cite{cooper_2010:van_waals,klimes_2010:chemical_accuracy,berland_2014:exchange_functional,hamada_2014:van_waals, Berland_2015:van_waals}.

In the GGA framework, e.g., PBE 
\cite{Perdew_1996:generalized_gradient}, the total GGA exchange can be
written as a functional of the electron density $n(\textbf{r})$ as
\begin{equation}\label{ggax}
E_{\rm x}^{\rm GGA}[n] = \int\;d\textbf{r}~n(\textbf{r})
\varepsilon_{\rm x}^{\rm hom}\big(n(\textbf{r})\big) F_{\rm x}(s)\;,
\end{equation}
where $\varepsilon_{\rm x}^{\rm hom}$ is the exchange energy of a
homogeneous electron gas, and $F_{\rm x}(s)$ is the enhancement factor,
a function of the reduced density gradient $s \propto |\nabla
n(\textbf{r})|/n(\textbf{r})^{4/3}$ that contributes to the total energy
of inhomogeneous systems. The precise form of $F_{\rm x}(s)$ varies from
functional to functional and its key features---such as its asymptotic
behavior at high $s$ or the maximum value of ${d}F_{\rm
x}(s)/{d}s$---have a direct and traceable impact on the energy and
forces of a system. In
Refs.~\cite{Jenkins_2021:reduced-gradient_analysis, Jenkins2024_Reduced}
we have investigated the ``signature'' of the reduced gradient for
several types of van der Waals complexes, effectively resolving the
interaction energy as a function of $s$. Through our reduced-gradient
analysis, we found that generally different classes of system tend to
have different $s$-signatures, meaning that individual classes of
systems can be targeted by an appropriately chosen enhancement factor to
yield greater accuracy.  For this reason, a suitable enhancement factor
should be flexible enough to target classes of systems with specific
reduced-gradient signatures, while not inducing spurious effects due to
overfitting.  While highly flexible generic functionals forms can be
constructed \cite{wellendorf_2012:density_functionals,
madsen_2007:functional_form}, we opted for a simple analytical form,
with high interpretability. 

The original vdW-DF3 functional variants, opt1 and opt2, used GGA
exchange forms inspired by B88 and
B86b~\cite{becke_1988:density_functional, becke_1986:large_gradient},
respectively. Both enhancement factors contained two adjustable
parameters: $\mu$, which controls the second derivative of $F_{\rm
x}(s)$ at low $s$, and $\kappa$, which primarily affects the ``tail'' of
$F_{\rm x}(s)$ at high $s$. In both of these functionals, and in vdW-DF3,
the $\mu$ value has been set to that of PBEsol
\cite{perdew_2008:restoring_density}. When paired with vdW-DF
correlation, functionals with values identical to this
\cite{klimes_2011:van_waals, hamada_2014:van_waals,
Chakraborty_2020:next-generation_nonlocal} or similar
\cite{cooper_2010:van_waals, berland_2014:exchange_functional} have been
found to provide accurate lattice constants for metals and covalent
solids \cite{tran_2019:nonlocal_van,
Chakraborty_2020:next-generation_nonlocal}.

\begin{figure}
\includegraphics[width=\columnwidth]{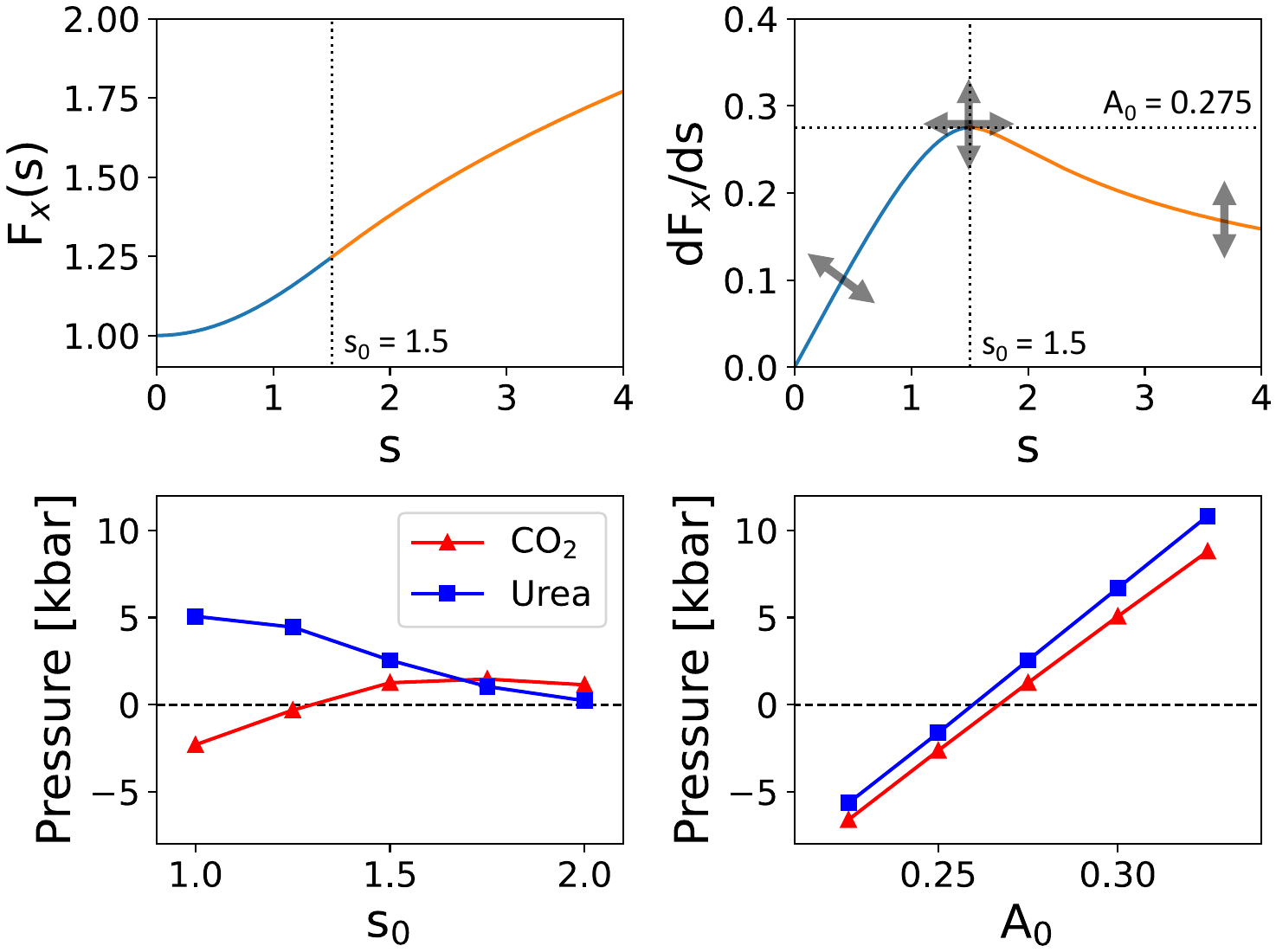}
\caption
{\label{fig:Fx_w3mc} \textbf{(top)} General form of the exchange enhancement factor
$F_{\rm x}(s)$ of vdW-DF3-mc and its derivative
with respect to $s$. The vertical and horizontal dotted lines indicate $s_0$ and $A_0$, respectively. Gray arrows indicate the degrees of freedom afforded by $s_0$, $A_0$, $\mu$ and $\kappa$. \textbf{(bottom)} As an example of the impact of $s_0$ and $A_0$ on physical quantities, we show their effect on the pressure for molecular crystals of CO$_2$ and urea in
their experimental unit cells.}
\end{figure}

To retain these degrees of freedom while including additional
flexibility for mid-range $s$, we define vdW-DF3-mc's enhancement
factor as
\begin{equation}\label{eq:df3-mc_Fx(s)}
F_{\rm x}(s) = \begin{cases} 1 + \mu
s^2 + As^4 + Bs^6,& \text{if } s < s_0\\ C + \kappa s^{2/5} + Ds^{-8/5}
+ Es^{-18/5},& \text{if } s \geq s_0.  \end{cases}
\end{equation}
Here, $\mu$ and $\kappa$ retain their functions as stated above. This
enhancement factor also implicitly includes parameters which we call
$s_0$ and $A_0$. We define $s_0$ as the value of $s$ at which ${
d}^2F_{\rm x}(s)/{d}s^2 = 0$, while $A_0$ is ${d}F_{\rm x}(s)/{d}s$
evaluated at $s_0$. In addition to setting $s_0$, $A_0$, and the
asymptotic coefficient $\kappa$, the coefficients $A$, $B$, $C$, $D$,
and $E$ are constrained to ensure that $F_{\rm x}(s)$ remains smooth to
2nd order and continuous at the boundary $s=s_0$.  The general form of
this exchange enhancement factor is plotted in Fig.~\ref{fig:Fx_w3mc},
showing the importance of the boundary at $s_0$ with respect to the
derivative $dF_{\rm x}/ds$. This figure also highlights the physical
importance of $s_0$ and $A_0$ with respect to their impact on system
geometries. Crystalline CO$_2$ and urea, which were among the molecular
crystals studied in Ref.~\cite{Jenkins_2021:reduced-gradient_analysis},
possess different signatures of the reduced density gradient. Using our
new form of $F_{\rm x}(s)$, these differences can be exploited to
simultaneously minimize errors in the unit cell geometry of both
systems.

%%%%%%%%%%%%%%%%%%%%%%%%%%%%%%%%%%%%%%%%%%%%%%%%%%%%%%%%%%%%%%%%%%%%%%%
\subsection{Re-optimizing Nonlocal Correlation}

In the van der Waals density functional of Dion \emph{et al.}
\cite{Dion_2004:van_waals}, the nonlocal correlation energy is expressed
as a functional of electron density $n(\textbf{r})$:
\begin{equation}\label{eq:nlc}
E_{\rm c}^{\rm nl}[n] = \frac{1}{2}
\int d\textbf{r}\;d\textbf{r}'n(\textbf{r})\,\Phi
(\textbf{r},\textbf{r}') \,n(\textbf{r}'),
\end{equation}
where $\Phi (\textbf{r},\textbf{r}')$ is a kernel to describe the
self-interaction of the electron density. By expanding the adiabatic
connection formula (ACF) to second order, it becomes possible to derive
such an expression for the kernel, where the nonlocal correlation energy
can be rewritten as
\begin{multline} \label{eq:nlc_expanded}
E_{\rm c}^{\rm nl}[n] =
\int_0^{\infty} \frac{du}{4\pi}\int \frac{{
d}\textbf{q}}{(2\pi)^3}\frac{{
d}\textbf{q}'}{(2\pi)^3}[1-(\hat{\textbf{q}}\cdot\hat{\textbf{q}}')^2]\times \\
S_{\textbf{q},\textbf{q}'}(iu)S_{\textbf{q}',\textbf{q}}(iu),
\end{multline}
where $\textbf{q}$ is a plasmon momentum, $u$ is an imaginary frequency, and the function
$S_{\textbf{q},\textbf{q}'}$ represents the plasmon propagator with
spectator contributions omitted. Partially derived from the dielectric
function of an electron gas, and made to obey four exact physical
constraints, the plasmon propagator takes the following form:
\begin{multline} \label{eq:Sqq}
S_{\textbf{q},\textbf{q}'}(\omega) =
\frac{1}{2}\int d\textbf{r}\: \omega_p^2(\textbf{r})\; e^{-i(\textbf{q}-\textbf{q}')\cdot \textbf{r}}\times \\ \bigg[\frac{1}{\big(\omega +
\omega_{q}(\textbf{r})\big)\big(-\omega + \omega_{q'}(\textbf{r})\big)} + \\
\frac{1}{\big(\omega +
\omega_{-q'}(\textbf{r})\big)\big(-\omega + \omega_{-q}(\textbf{r})\big)}\bigg].
\end{multline}
Here, $\omega = iu$, while 
$\omega_p(\textbf{r})$ is the classical plasmon
frequency $\sqrt{4\pi n(\textbf{r})}$, and $\omega_q(\textbf{r})$ is an
appropriately chosen plasmon dispersion law. To allow for effective parameterization of the nonlocal kernel,
a single length-scale $1/q_0(\textbf{r})$ is used for the plasmon dispersion, so that 
\begin{equation} \label{eq:disp_law} \omega_q(\textbf{r}) =
\frac{q^2}{2}\frac{1}{h(q/q_0(\textbf{r}))}, \end{equation}
where $h(y)$ is a switching function which constrains the asymptotic
behavior of $\omega_q$. At high $q$, $\omega_q$ behaves as $q^2/2$ to
reproduce the exactly-known self-correlation, while at low $q$ it
becomes constant.  The value of $q_0(\textbf{r})$ is chosen so that a
first-order expansion of the ACF in $S$ yields a GGA-type form of the
local exchange-correlation functional. This so-called ``internal
functional'' $\varepsilon_{\rm xc}^{\rm int}$, can be written as
\begin{equation}\label{eq:e_xc^int} \begin{split} \varepsilon_{\rm
xc}^{\rm int} & = \pi \int \frac{{
d}\textbf{q}}{(2\pi)^3}\bigg[\frac{1}{\omega_q(\textbf{r})} -
\frac{2}{q^2}\bigg] \\ & = -\frac{1}{\pi} q_0(\textbf{r})
\int_0^{\infty} dy[1-h(y)].  \end{split} \end{equation}
In vdW-DF1 (and vdW-DF2), the switching function is chosen as $h_{\rm DF1}(y) = 1 -
e^{-4\pi y^2/9}$ so that the above integral over $y$ evaluates to $3/4$.
With this, $q_0(\textbf{r})$ can be conveniently written as a ratio of
the internal and LDA exchange functionals:
\begin{equation}\label{eq:q0} q_0(\textbf{r}) = (\varepsilon_{\rm
xc}^{\rm int}/\varepsilon_{\rm x}^{\rm LDA})k_{\rm F}(\textbf{r}),
\end{equation}
where $k_{\rm F}(\textbf{r})$ is the Fermi wavevector, equal to $(3\pi^2
n(\textbf{r}))^{1/3}$. In the more recent development of vdW-DF3,
this definition of $q_0(\textbf{r})$ is retained, but the switching
function is chosen as
\begin{equation} \label{eq:h_DF3} h_{\rm DF3}(y) =  1 -
\frac{1}{1+\gamma y^2 + (\gamma^2 - \beta)y^4 + \alpha y^8},
\end{equation}
where $\gamma$, $\beta$, and $\alpha$ are variational parameters. With
this definition of $h_{\rm DF3}(y)$, the plasmon frequency can be
expanded as
\begin{equation} \label{eq:omega_DF3} \omega_q \sim \frac{y^2}{h_{\rm
DF3}(y)} = \frac{1}{\gamma} + \frac{\beta}{\gamma^2}y^2 +
\bigg(\frac{\beta^2}{\gamma^3}- \frac{2\beta}{\gamma} + \gamma\bigg)y^4
+ \dots  \end{equation}
The low-$q$ behavior of $\omega_q$ in vdW-DF3 should have been well
accounted-for by $\gamma$ and $\beta$, with $\alpha$ serving to
normalize Eq.~(\ref{eq:e_xc^int}).  However, we have found that in
practice, the constraint on $\alpha$ counters the intended effects of
the $\gamma$ parameter. As a result, $h_{\rm DF3}(y)$ offered less
flexibility than anticipated, as $E_{\rm c}^{\rm nl}$ remained somewhat
insensitive to the choice of parameters.

To amend this and restore flexibility to $h_{\rm DF3}(y)$ in vdW-DF3-mc,
we instead constrain $\alpha$ for smoothness, rather than fixing it
exactly to reproducing the exchange from Eq.~(\ref{eq:q0}).  For given
values of $\gamma$ and $\beta$, $\alpha$ is chosen to minimize the mean
curvature of $h(y)$, i.e., the finite integral over $d^2h(y)/dy^2$.
While this choice results in a $q_0(\textbf{r})$ that is some scalar
factor different from that of vdW-DF1, it has no significant impact on
the evaluation of the kernel $\Phi(\textbf{r},\textbf{r}')$, nor does it
break any of the four exact physical constraints of vdW-DF1's original
design. One may also find precedent for such a design choice in the
so-called vdW-DF09 by Vydrov and van Voorhis
\cite{Vydrov_2009:improving_accuracy}, which also broke the constraint
on Eq.~(\ref{eq:q0}) in addition to using a different form of
$S_{\textbf{q},\textbf{q}'}$.

%%%%%%%%%%%%%%%%%%%%%%%%%%%%%%%%%%%%%%%%%%%%%%%%%%%%%%%%%%%%%%%%%%%%%%%
\subsection{Optimization Set and Procedure}\label{subsec:opt}

\begin{figure}
\includegraphics[width=\columnwidth]{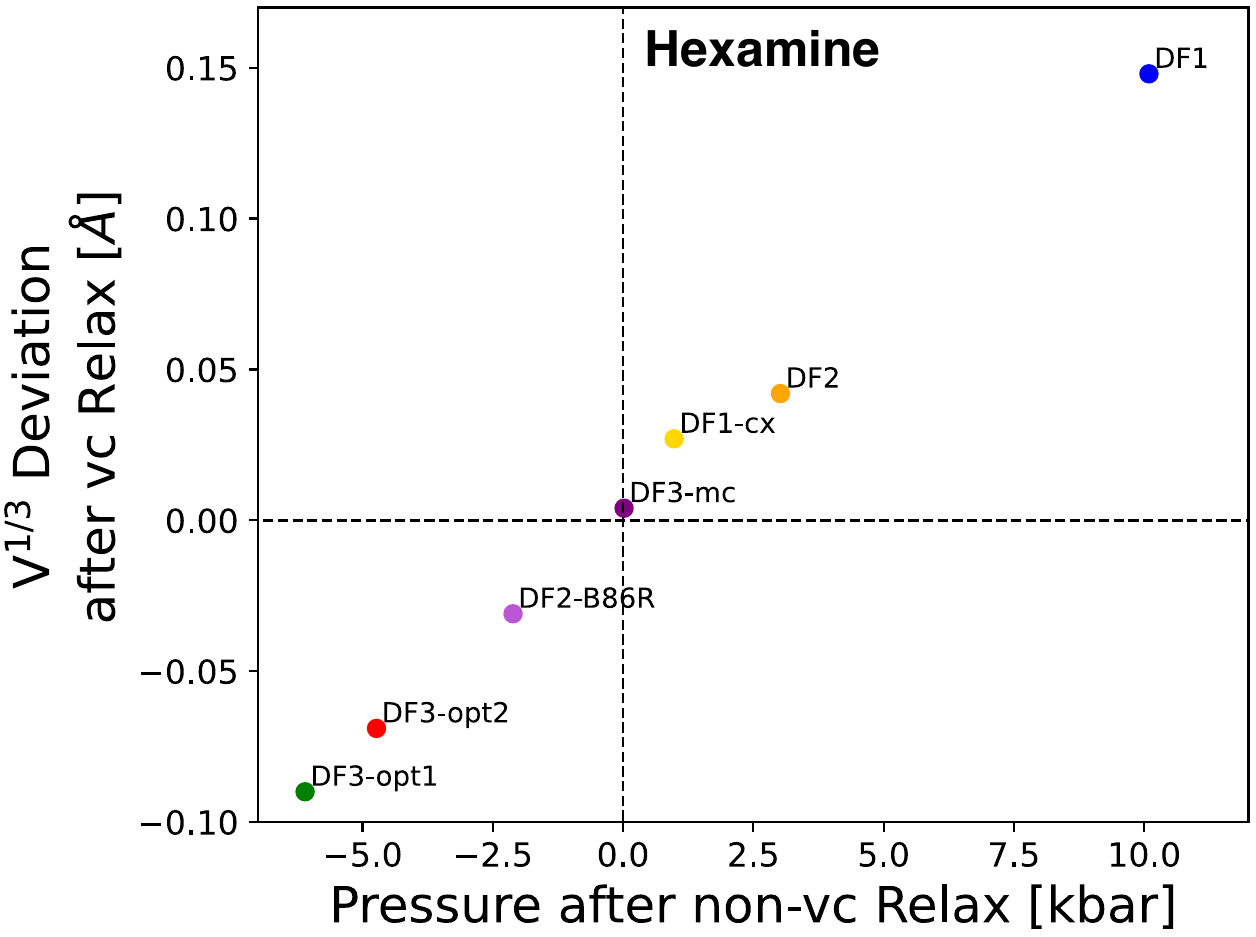}
\caption {\label{fig:hexamine} Comparison of structure residuals for
several vdW-DF variants on hexamine. On the horizontal axis, a
non-variable cell relaxation is performed on the experimental unit cell
reported in Ref.~\cite{moellmann_2014:dft-d3_study}, and the resulting
pressure is reported. On the vertical axis, a full variable cell
relaxation is performed, and the resulting deviation from experimental
volume is given.}
\end{figure}

Our optimization set for vdW-DF3-mc was comprised of three parts: the
full X23 set of molecular crystals, the two layered systems graphite and
hexagonal boron nitride ($h$-BN), and 24 hand-selected molecular dimers.
Fourteen of the dimers are from the S22$\times$5, including four
H-bonded dimers, prioritizing those with multiple bonds, and five each
from the dispersion and mixed-character groups, selected for
representation of hydrocarbon interactions and $\pi$--$\pi$ stacking.
The remaining ten dimers come from the S66$\times$8, with similar
criteria. Four of these systems are H-bonded, including the water dimer,
while the other six are $\pi$--$\pi$ and TS configurations of benzene
and nucleotide dimers. In cases where a particular dimer was present in
both sets, we used the S66$\times$8 configurations as our reference due
to its larger sampling of non-equilibrium geometries. The full list of
systems used in vdW-DF3-mc's optimization, along with their data sets of
origin, is given in the Supporting Materials.

With the inclusion of layered systems, we improve performance for more
two-dimensional molecular crystals.  Similarly, dimers were included to
better model long-range interactions, which are especially prevalent in
porous materials such as COFs and HOFs.  Note that the use of X23
molecular crystals as a reference set, without back-correction for
zero-point and thermal expansion, contains reference systems with
geometries reflecting different temperatures.  Thus, vdW-DF3-mc aims to
be a pragmatic functional for practical molecular crystal modeling, and
ultimately its utility depends on its performance.  An alternative
choice would be to back-correct for thermal and zero-point corrections,
i.e., following the methodology of
Ref.~\cite{dolgonos_2019:revised_values}.  However, such back-correction
is dependent on the functional, and also requires costly phonon
calculations at different lattice constants for comparison with
experiment. 

In our optimization, we began by optimizing solely the exchange,
calculating the stress tensors of our test systems in their reference
geometries. As discussed earlier, we set $\mu = \mu_{\rm PBEsol} \approx
0.1234$, and optimize with respect to $s_0$, $A_0$, and $B$ of
Eq.~(\ref{eq:df3-mc_Fx(s)}).  To do so, we performed non-variable cell
relaxations of each system in their experimentally-measured unit cell
shape/volume.  Not only is this method less costly than using variable
cell relaxations---an important consideration for the early stages of
optimization---but as Fig.~\ref{fig:hexamine} shows for the case of
the hexamine molecular crystals, the cell pressure resulting from a
given functional is strongly correlated with the volume deviation
after variable cell relaxation.  Rather than the pressure, which only
averages diagonal elements of the stress tensor, we minimized the mean
deviation (MD) of all stress tensor elements for all 3D structures in
our test set. That is, for the $n$ systems in our set, we optimize
with respect to
\begin{equation} {\rm MD} = \frac{1}{n}\sum_{i=1}^n
\sigma_{xx}^i+\sigma_{yy}^i+\sigma_{zz}^i +
2(\sigma_{xy}^i+\sigma_{xz}^i+\sigma_{yz}^i), \end{equation}
%i
where $\sigma_{jk}^i$ is a stress tensor element of the system indexed
by $i$. By including the off-diagonal elements, $\sigma_{xy}$, $\sigma_{xz}$, and $\sigma_{yz}$, our optimization limits errors in the shear modulus.

\begin{figure}
\includegraphics[width=\columnwidth]{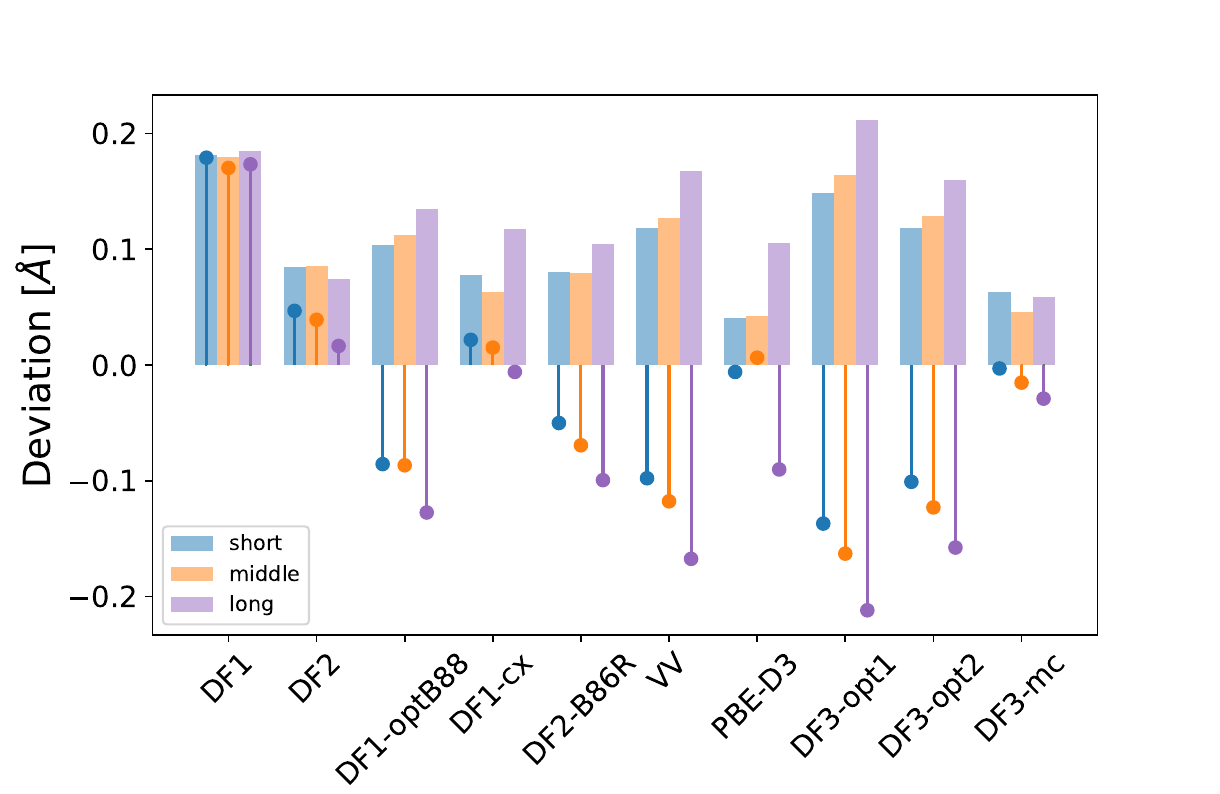}
\caption[Accuracy of DF3-mc for the X23 molecular
crystal axis lengths] {\label{fig:X23_axes} Average deviations (lines)
and absolute deviations (bars) in X23 cell axis lengths for several
different functionals. Axes are subdivided into the shortest (blue),
middle (orange), and longest (purple) for each given system.}
\end{figure}

Another benefit of stress-based optimization can be seen in
Fig.~\ref{fig:X23_axes}, which shows the deviation in individual cell
axis lengths within the X23 set. Because the X23 was used in the
optimization of vdW-DF3-mc, it is perhaps not surprising that its
overall deviation from experiment is low.  But it is still interesting
to note that cell axes of any size deviate by approximately the same
amount. We speculate that this is because optimizing with the stress
tensor as an objective function rewards accurate estimation of cell
shape, and not just volume. This approach may have also improved
accuracy on hydrogen-bonded frameworks like acetic acid, where large
errors in the stress (due to stronger magnitude of the forces involved
in hydrogen-bonding) could be hidden as smaller errors in cell
geometries.

Next, taking the optimized exchange, we calculated binding energies for
our test set with varying combinations of $\gamma$ and $\beta$ from
Eq.~(\ref{eq:h_DF3}). For molecular crystals and layered structures, we
calculated the cohesive energy per fragment as
\begin{equation}\label{Eb_crys} E_{\rm coh} = \frac{1}{N}E_{\rm tot} -
\frac{1}{N}\sum_{i}E_i, \end{equation}
for a system comprised of $N$ separate molecules/layers. For the dimers
in our test set, we calculate the total binding energy as
\begin{equation}\label{Eb_dimer} E_{\rm bind} = E_{\rm tot} - E_A - E_B,
\end{equation}
with $E_A$ and $E_B$ being the energies of molecule $A$ and $B$ in the
gas phase, respectively. We optimize our nonlocal correlation functional
with respect to deviations from experimental cohesive energies; or, in
the case of dimers with different separations, deviations from the
CCSD(T) calculated binding energies
\cite{grafova_2010:comparative_study, rezac_2011:s66_well-balanced}. We
simultaneously minimize the mean absolute relative deviation (MARD) of
the cohesive energy for the solids, and the weighted mean absolute
relative deviation (WMARD) for the dimers, which accounts for the
smaller interaction energies of non-equilibrium geometries. These are
defined as
\begin{align}\label{eq:MARD,WMARD} {\rm MARD} & = \frac{1}{n}\sum_{{\rm
sys}=i}^n \frac{|E_{\rm sys}^{\rm DFT} - E_{\rm sys}^{\rm ref}|}{E_{\rm
sys}^{\rm ref}}\times 100,\\
{\rm WMARD} & =
\frac{1}{n}\frac{1}{m}\sum_{{\rm sys}=1}^n \sum_{{\rm sep}=1}^m
\frac{|E_{\rm sys,sep}^{\rm DFT} - E_{\rm sys,sep}^{\rm ref}|}{E_{\rm
sys,opt}^{\rm ref}}\times 100.  \end{align}
For the dimers, $E_{\rm sys,opt}^{\rm ref}$ denotes the
reference energy of the dimer at equilibrium distance.

When optimizing nonlocal correlation with respect to the energy, weights
are applied to the deviation of each system depending on its type.
Molecular crystals were weighed to make up 50\% of the set's MARD, while
the layered systems and dimers made up the remaining 5\% and 45\%,
respectively. We further emphasized the importance of hydrogen bonding
within the dimers, weighing those systems twice as much as the
dispersion-dominated and mixed-character dimers.

\begin{table}
\caption
{\label{tab:params} Complete list of parameters for vdW-DF3-mc, compared with vdW-DF3-opt1 and -opt2.
The parameters varied/optimized in this study were $s_0$, $A_0$, $\kappa$, $\alpha$, $\beta$, and $\gamma$, although
$\beta$ came out to be zero. The internal
parameters are functions of (and completely determined by) the external parameters.}
\begin{tabular*}{\columnwidth}{@{\extracolsep{\fill}}lccc@{}}
\hline\hline & vdW-DF3-mc & vdW-DF3-opt1 & vdW-DF3-opt2 \\ \hline
\multicolumn{4}{c}{\emph{exchange (external)}}\\
$\mu$ & $10/81$ & $10/81$ & $10/81$ \\
$s_0$ & 1.50 & 3.14 & 1.51 \\
$A_0$ & 0.275 & 0.208 & 0.201 \\
$\kappa$ & 0.880 & --- & 0.426 \\[2ex]
\multicolumn{4}{c}{\emph{exchange (dependent)}}\\
$A$ & $-3.944\times 10^{-3}$ & --- & --- \\
$B$ & $-9.246\times 10^{-4}$ & --- & --- \\
$C$ & 0.2539 & --- & --- \\
$D$ & $-0.1444$ & --- & --- \\
$E$ & $0.1462$ & --- & --- \\[2ex]
\multicolumn{4}{c}{\emph{nonlocal correlation}}\\
$Z_{ab}$ & $-1.887$ & $-0.8491$ & $-1.887$ \\
$\alpha$ & 0.0532 & 0.9495 & 0.2825 \\
$\beta$ & 0.0 & 0.0 & 0.0 \\
$\gamma$ & 1.42 & 1.12 & 1.29 \\[1ex] \hline\hline
\end{tabular*} \end{table}

Once we were sufficiently close to the optima for exchange and
correlation, we performed a final search of the four parameters $A_0$,
$s_0$, $\kappa$, and $\gamma$. We optimized with respect to both binding
energies and unit-cell geometries, the latter of which were calculated
via variable cell relaxations of the molecular crystals and layered
structures. Doing so yielded the parameters for the GGA exchange: $A_0 =
0.275$, $s_0 = 1.50$, and $\kappa = 0.88$, and for the nonlocal
correlation: $\alpha = 0.0532$, $\beta = 0.0$, $\gamma = 1.42$. These
values, along with the equivalent values of vdW-DF3-opt1 and -opt2, are
listed in Table \ref{tab:params}. In this table, we also list, for
convenience, the explicit dependent exchange parameters used in
Eq.~(\ref{eq:df3-mc_Fx(s)}). Note that opt1, with  its B88 exchange
form, lacks an $s^{2/5}$  asymptote and thus has no  equivalent to
$\kappa$. Also listed in Table \ref{tab:params} are the parameters of
our nonlocal correlation functional, which includes the internal
functional's gradient correction coefficient $Z_{ab}$. In vdW-DF3-opt1,
the nonlocal correlation's gradient correction coefficient $Z_{ab}$ is
taken to be the same as in vdW-DF1, while both vdW-DF3-opt2 and -mc take
the $Z_{ab}$ value of vdW-DF2.

\begin{figure}
\includegraphics[width=\columnwidth]{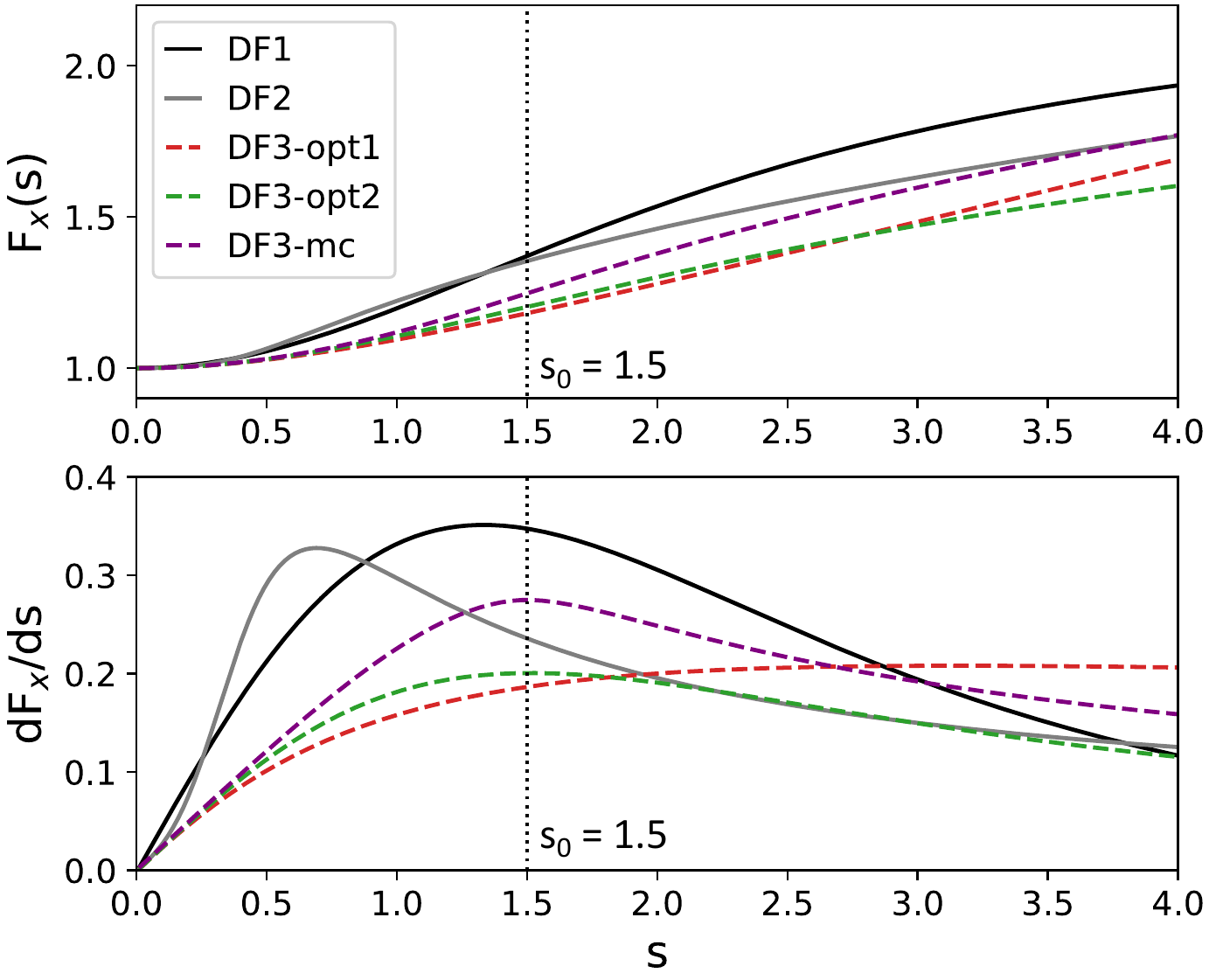}
\caption {\label{fig:Fx_DF3-mc} \textbf{(top)} The exchange enhancement
factor $F_{\rm x}(s)$ of vdW-DF3-mc and \textbf{(bottom)} the derivative
of $F_{\rm x}$ with respect to $s$, compared with those of several other
vdW-DF variants (revPBE for vdW-DF1, PW86r for vdW-DF2, W31X and W32X for
vdW-DF3-opt1 and -opt2, respectively). The vertical dotted line
indicates the location of $s_0$, the boundary between the two parts of
Eq.~(\ref{eq:df3-mc_Fx(s)}).}
\end{figure}

\begin{figure}
\includegraphics[width=\columnwidth]{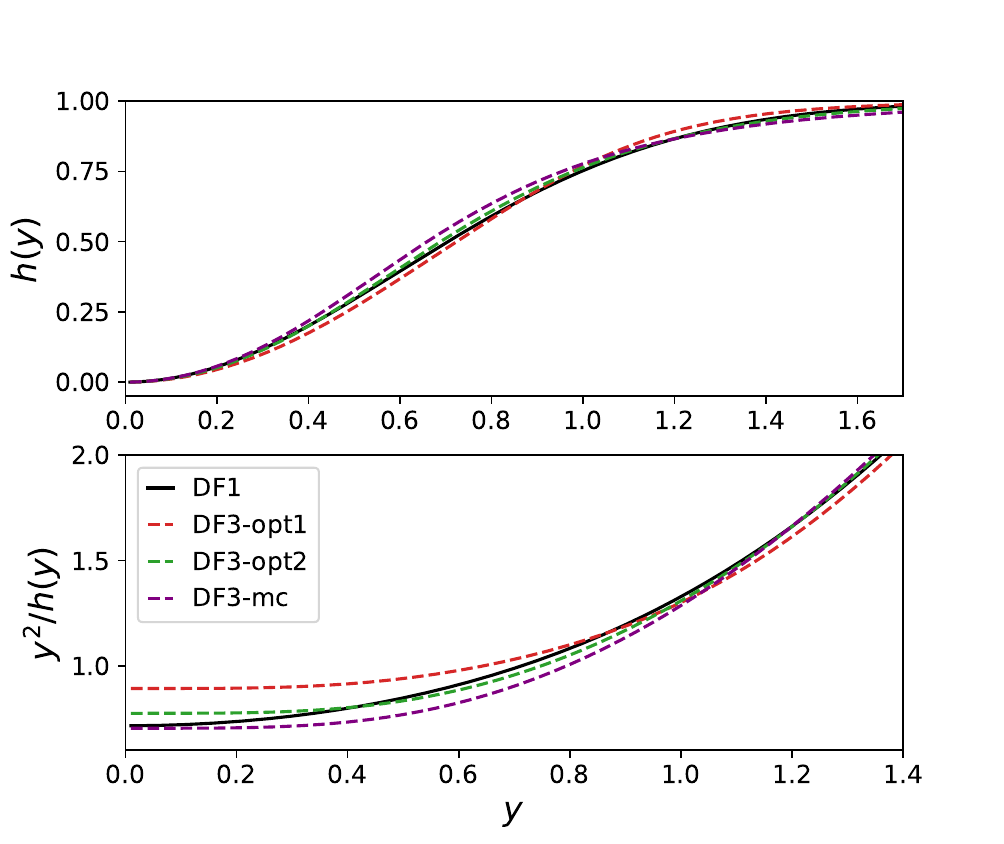}
\caption
{\label{fig:hy_DF3-mc} \textbf{(top)} The switching function $h(y)$ of
vdW-DF3-mc's nonlocal correlation term, and \textbf{(bottom)} its
corresponding $y^2/h(y)$ function, which is analogous to the
plasmon dispersion law $\omega_q$. Both functions are compared with
those of vdW-DF1, and the previous two optimizations of vdW-DF3.}
\end{figure}

The exchange enhancement factor of vdW-DF3-mc and its derivative is
compared with those of several other vdW-DF's in
Fig.~\ref{fig:Fx_DF3-mc}. Here, we observe that our peak in ${d}F_{\rm
x}(s)/{d}s$ occurs at higher $s$ than than of either vdW-DF1 or vdW-DF2,
ensuring more slowly-varying corrections due to the density gradient. On
the other hand, vdW-DF3-mc generates a larger contribution than either
of its predecessors opt1 or opt2, particularly in the region
$1.5<s<3.5$. This region was found to be the largest contributor of
repulsion in the X23 molecular crystals, a positive indicator that this
optimization avoids the overbinding of vdW-DF3-opt1 and -opt2
\cite{Jenkins_2021:reduced-gradient_analysis,
Chakraborty_2020:next-generation_nonlocal}.

Lastly, we show in Fig.~\ref{fig:hy_DF3-mc} the switching function of
vdW-DF3-mc's nonlocal correlation, alongside other known switching
functions for comparison. Despite the differing constraints on
$q_0(\textbf{r})$ between these functionals, we note no significant
visual distinctions in their plasmon dispersions. It is also intriguing
to note that, similar to the optimizations of vdW-DF3-opt1 and -opt2, we
find an optimum $\beta$ very near zero. Moreover, the energy was found
to be relatively insensitive to our choice in $\beta$. This is in
contrast to $\gamma$, which now more effectively changes the strength of
the nonlocal correlation contributions, after altering the constraint on
$\alpha$. For this reason, and for consistency with the two earlier
vdW-DF3 forms, we have set $\beta=0$.

%%%%%%%%%%%%%%%%%%%%%%%%%%%%%%%%%%%%%%%%%%%%%%%%%%%%%%%%%%%%%%%%%%%%%%%
\section{Computational Details}\label{vdw-df3-mc:computational_details}
%%%%%%%%%%%%%%%%%%%%%%%%%%%%%%%%%%%%%%%%%%%%%%%%%%%%%%%%%%%%%%%%%%%%%%%

All calculations were done using the \textsc{quantum espresso}
package \cite{giannozzi_2017:advanced_capabilities}. We used the SG15
optimized norm-conserving Vanderbilt (ONCV) pseudopotentials
\cite{Hamann_2013:optimized_norm-conserving}.  For optimizations of the
GGA exchange and nonlocal correlation, we find convergence for a
wavefunction cutoff of 80 Rydberg and a charge density cutoff of 320 Rydberg.
Energies and forces in the system were made to converge within $1 \times
10^{-6}$ Rydberg and $1 \times 10^{-4}$ Rydberg/bohr respectively. For
variable cell structural optimizations, a convergence threshold of 0.5
kbar was applied with respect to the unit-cell pressure.  For initial
searches in the exchange parameter space, we performed non-variable cell
relaxations on the 3D materials in our test set, allowing only atomic
positions to optimize. For the final optimization of both the exchange
and correlation, we do full variable cell relaxations. For our
optimization, we minimize the deviation from reference values in
Refs.~\cite{moellmann_2014:dft-d3_study, grafova_2010:comparative_study,
rezac_2011:s66_well-balanced} with respect to unit cell
stress/dimensions and binding energies.  We perform additional
calculations, testing the optimized vdW-DF3-mc on several validation
sets of molecular solids and layered structures, including several
well-studied polymorphic materials. These calculations use the same
convergence criteria as in the functional optimization.

In the following section, we compare the benchmark performance of
vdW-DF3-mc (hereafter abbreviated as DF3-mc) to that of several other
nonlocal and dispersion-corrected functionals.  In particular, we
examine the functionals vdW-DF \cite{Dion_2004:van_waals}, vdW-DF-optB88
\cite{klimes_2010:chemical_accuracy}, vdW-DF-cx
\cite{berland_2014:exchange_functional}, vdW-DF2
\cite{Lee_2010:higher-accuracy_van}, vdW-DF2-B86R
\cite{hamada_2014:van_waals}, vdW-DF3-opt1
\cite{Chakraborty_2020:next-generation_nonlocal}, and vdW-DF3-opt2
\cite{Chakraborty_2020:next-generation_nonlocal}. For brevity, they are
referred to as DF1, DF1-optB88, DF1-cx, DF2, DF2-B86R, DF3-opt1, and
DF3-opt2, respectively. For the X23 molecular crystals, data for the
rVV10 nonlocal functional \cite{Sabatini_2013:nonlocal_van} is reported
by us in Ref.~\cite{Chakraborty_2020:next-generation_nonlocal} and used
in this paper for comparison, denoted as VV.

%%%%%%%%%%%%%%%%%%%%%%%%%%%%%%%%%%%%%%%%%%%%%%%%%%%%%%%%%%%%%%%%%%%%%%%
\section{Results}\label{vdw-df3-mc:results}
%%%%%%%%%%%%%%%%%%%%%%%%%%%%%%%%%%%%%%%%%%%%%%%%%%%%%%%%%%%%%%%%%%%%%%%

%%%%%%%%%%%%%%%%%%%%%%%%%%%%%%%%%%%%%%%%%%%%%%%%%%%%%%%%%%%%%%%%%%%%%%%
\subsection{Studied Benchmark Sets}

Section~\ref{subsec:opt} outlines the sets used to optimize vdW-DF3-mc,
including molecular crystals from the X23 set, and dimers from the S22$\times$5
and S66$\times$8. In addition to these systems, we use several other benchmark
sets to test the accuracy, which are described here in detail.

We make use of the aforementioned X23 set of molecular crystals in its
original form \cite{moellmann_2014:dft-d3_study}, compiled by Moellmann
and Grimme as an extension of the C21 set compiled by Otero-de-la-Roza
\emph{et al.} \cite{otero_2012:benchmark_non-covalent}. This version of
the X23 uses zero-point vibrational corrections to the experimental
sublimation enthalpy. Another benchmark that we use is the G60, a larger
and somewhat more diverse molecular crystal set that also includes some
halogenated materials. Originally compiled by Maschio \emph{et al.}\ in
Ref.~\cite{Maschio_2011:intermolecular_interaction}, we use the
reference energies of Ref.~\cite{Cutini_2016:assessment_different},
which applies a constant $-2RT$ correction to experimental sublimation
enthalpies. Experimental structures for the G60 systems were extracted
from the Cambridge Structural Database
(CSD)~\cite{groom_2016:cambridge_structural} for comparison with our
functional. CSD reference codes for each system are given in the
Supporting Materials. We also note some difficulty in the convergence of
1,3,5-trinitrobenzene.  For this reason, it is omitted from our
summaries of the G60 data.

We also examine polymorphs of ice, some of which are compiled in the
ICE10 and DMC-ICE13 data sets. ICE10 compiles ten ice polymorphs with
lattice parameters drawn from low-temperature neutron diffraction
experiments \cite{Brandenburg_2015:benchmarking_dft}. The DMC-ICE13 set
provides lattice energies for each of the ICE10 polymorphs, plus ice IV,
XI, and XVII, which are taken from diffusion Monte Carlo (DMC)
calculations \cite{DellaPia_2022:dmc-ice13_ambient}.

The POLY59 set compiles a total of 59 polymorphs of five different
molecular solids \cite{Brandenburg_2016:organic_crystal}, based on the
sixth blind test of organic crystal structure prediction
\cite{Reilly_2016:report_sixth}.  Each of the five compounds possesses
one or more experimentally-known configurations, which serve as
estimates for the configurational ground state. The ability to compare
other polymorphs with these structures thus makes the POLY59 useful as a
blind test of different computational methods.

Beyond the scope of 3D molecular solids, we also examine several
hand-picked 2D layered systems. For ease of comparison, we take the same
nine layered systems compiled for benchmarking in the development of
DF3-opt1 and -opt2 \cite{Chakraborty_2020:next-generation_nonlocal}. Reference data for interlayer separations and cohesive energies are taken from experiment and random phase approximation (RPA) calculations, respectively \cite{Bjorkman_2012:vdw_bonding_layered,Bjorkman_2014:testing_several}. We
also study several molecular dimers from the range-separated
S22$\times$5 and S66$\times$8 data sets
\cite{grafova_2010:comparative_study, rezac_2011:s66_well-balanced}.

Statistical data for the G60, X23, and layered structures is summarized
in Table~\ref{tab:df3-mc_all}, with comparisons to several other van der
Waals functionals and PBE-D3. Raw data for all relevant calculations is
provided in the Supporting Materials, including the S66$\times$8 and
S22$\times$5 dimers.  That said, we emphasize that DF3-mc is,
first-and-foremost, designed for the accurate description of molecular
solids. The inclusion of dimers is thus intended as a stand-in for
certain long-range bond characters, and not to imply more general-use
cases.

\begin{table*}\centering
\caption[Statistical summary of several studied benchmark sets]
{\label{tab:df3-mc_all} Comparison of various statistical measures for
cohesive energy per molecule and structure parameters in the X23, G60,
and layered structure benchmark sets. Structural accuracy is measured
for the cube root of the volume in the X23 and G60, and the inter-layer
separation in layered structures. All data for individual systems is
available in the Supporting Materials.}
\begin{tabular*}{\textwidth}{@{\extracolsep{\fill}}lrrrrrrrr@{}}
\hline\hline & DF1 & DF2 & DF1-cx & DF2-B86R & PBE-D3 & DF3-opt1 & DF3-opt2 & DF3-mc \\ \hline
\multicolumn{9}{c}{\bf Structure Parameters}\\
\multicolumn{9}{c}{\emph{vdW-bonded X23 (10)}}\\
MD [\AA]  & 0.176 & 0.019 & 0.035 & $-0.072$ & $-0.017$ & $-0.169$ & $-0.132$ & $-0.023$ \\
MAD [\AA] & 0.176 & 0.048 & 0.051 & 0.072 & 0.030 & 0.169 & 0.132 & 0.031 \\
MARD [\%] & 2.54 & 0.69 & 0.77 & 0.99 & 0.43 & 2.37 & 1.85 & 0.42 \\
\multicolumn{9}{c}{\emph{H-bonded X23 (13)}}\\
MD [\AA]  & 0.176 & 0.045 & $-0.010$ & $-0.075$ & $-0.018$ & $-0.175$ & $-0.128$ & $-0.010$ \\
MAD [\AA] & 0.176 & 0.053 & 0.042 & 0.075 & 0.028 & 0.175 & 0.128 & 0.029 \\
MARD [\%] & 2.60 & 0.75 & 0.64 & 1.18 & 0.44 & 2.70 & 1.98 & 0.45 \\
\multicolumn{9}{c}{\emph{All X23 (23)}}\\
MD [\AA]  & 0.176 & 0.034 & 0.010 & $-0.074$ & $-0.017$ & $-0.172$ & $-0.130$ & $-0.016$ \\
MAD [\AA] & 0.176 & 0.051 & 0.046 & 0.074 & 0.029 & 0.172 & 0.130 & 0.030 \\
MARD [\%] & 2.57 & 0.72 & 0.70 & 1.10 & 0.44 & 2.56 & 1.92 & 0.44 \\
\multicolumn{9}{c}{\emph{G60 (59)}}\\
MD [\AA]  & 0.095 & $-0.037$ & $-0.063$ & $-0.151$ & $-0.098$ & $-0.255$ & $-0.215$ & $-0.095$ \\
MAD [\AA] & 0.097 & 0.048 & 0.064 & 0.151 & 0.098 & 0.255 & 0.215 & 0.095  \\
MARD [\%] & 1.15 & 0.60 & 0.77 & 1.83 & 1.21 & 3.08 & 2.59 & 1.14 \\
\multicolumn{9}{c}{\emph{Layered (9)}}\\
MD [\AA]  & 0.38 & 0.41 & $-0.04$ & 0.03 & $-0.05$ & $-0.02$ & 0.00 & 0.06 \\
MAD [\AA] & 0.38 & 0.41 & 0.06 & 0.07 & 0.07 & 0.05 & 0.06 & 0.09 \\
MARD [\%] & 6.81 & 7.24 & 1.32 & 1.36 & 1.26 & 1.14 & 1.33 & 1.75 \\[3ex]

\multicolumn{9}{c}{\bf Cohesive Energy}\\
\multicolumn{9}{c}{\emph{vdW-bonded X23 (10)}}\\
MD [eV]  & $-0.083$ & $-0.051$ & $-0.074$ & 0.002 & 0.007 & $-0.097$ & $-0.088$ & $-0.017$ \\
MAD [eV] & 0.083 & 0.065 & 0.074 & 0.036 & 0.031 & 0.097 & 0.088 & 0.032 \\
MARD [\%] & 12.81 & 9.84 & 10.43 & 4.46 & 4.58 & 13.84 & 12.51 & 4.18 \\
\multicolumn{9}{c}{\emph{H-bonded X23 (13)}}\\
MD [eV]  & 0.002 & $-0.019$ & $-0.082$ & $-0.035$ & $-0.034$ & $-0.177$ & $-0.127$ & 0.002 \\
MAD [eV] & 0.042 & 0.048 & 0.088 & 0.052 & 0.054 & 0.177 & 0.127 & 0.029 \\
MARD [\%] & 3.74 & 4.73 & 9.33 & 5.32 & 6.06 & 19.34 & 13.99 & 2.61 \\
\multicolumn{9}{c}{\emph{All X23 (23)}}\\
MD [eV]  & $-0.035$ & $-0.033$ & $-0.079$ & $-0.019$ & $-0.016$ & $-0.142$ & $-0.110$ & $-0.006$ \\
MAD [eV] & 0.060 & 0.055 & 0.082 & 0.045 & 0.044 & 0.142 & 0.110 & 0.030 \\
MARD [\%] & 7.68 & 6.95 & 9.81 & 4.95 & 5.42 & 16.95 & 13.34 & 3.29 \\
\multicolumn{9}{c}{\emph{G60 (59)}}\\
MD [eV]  & $-0.153$ & $-0.123$ & $-0.173$ & $-0.061$ & 0.002 & $-0.208$ & $-0.179$ & $-0.065$ \\
MAD [eV] & 0.157 & 0.125 & 0.173 & 0.074 & 0.087 & 0.208 & 0.179 & 0.078 \\
MARD [\%] & 15.28 & 12.18 & 16.65 & 7.24 & 7.71 & 20.19 & 17.47 & 7.59 \\
\multicolumn{9}{c}{\emph{Layered (9)}}\\
MD [meV/\AA$^2$]  & $-2.31$ & $-2.16$ & 6.01 & 3.66 & 10.91 & 2.71 & 4.61 & 4.75 \\
MAD [meV/\AA$^2$] & 3.84 & 3.50 & 6.01 & 3.66 & 10.91 & 2.71 & 4.61 & 4.75 \\
MARD [\%] & 16.08 & 13.54 & 30.34 & 19.60 & 51.70 & 13.56 & 24.38 & 26.99 \\[1ex] \hline\hline
\end{tabular*} \end{table*}

\begin{figure}
\includegraphics[width=\columnwidth]{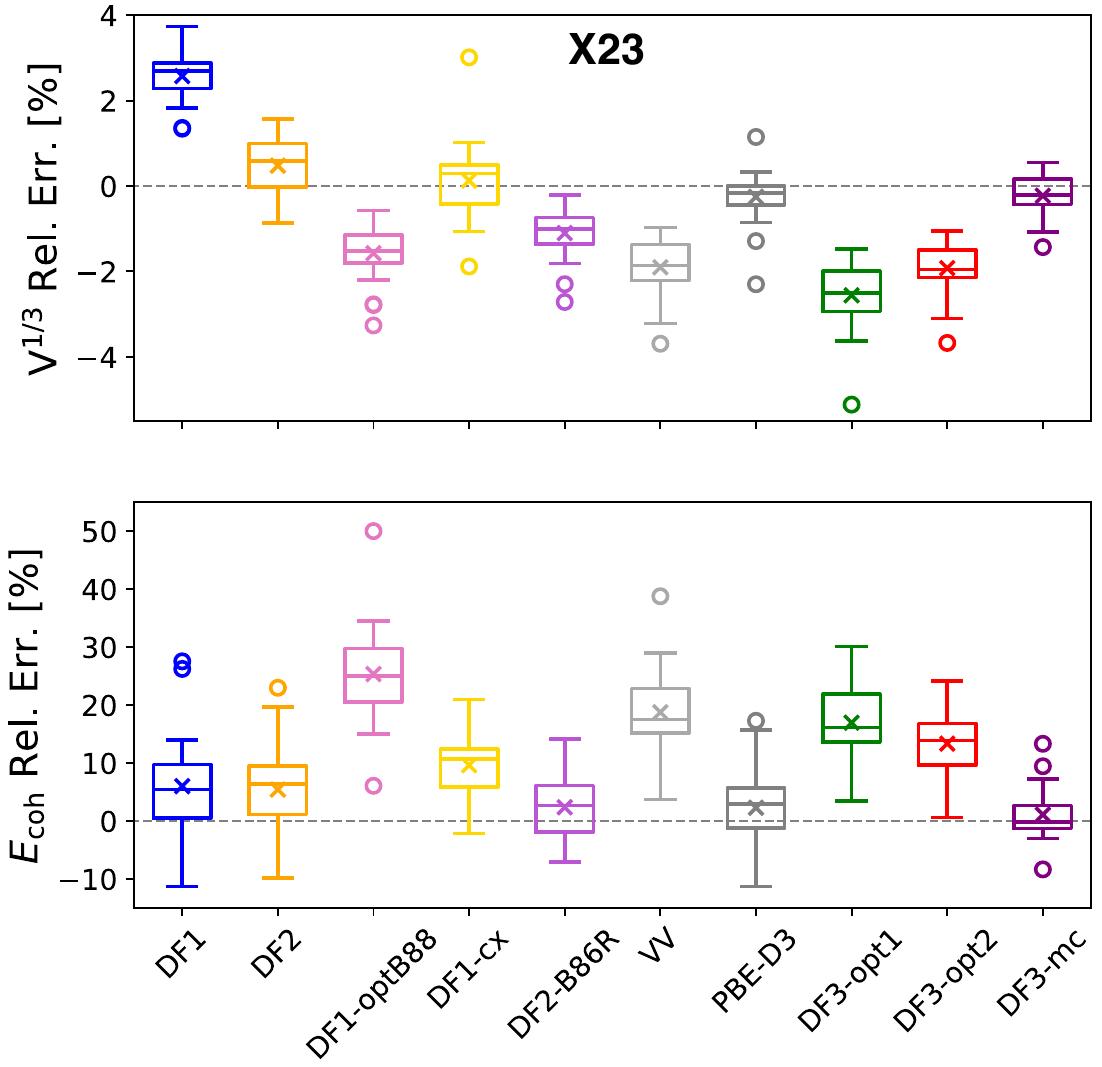}
\caption[Accuracy of DF3-mc for the X23 molecular
crystals] {\label{fig:X23_boxes} Accuracy of DF3-mc on the X23 set of
molecular solids, comparing relative deviations in cube roots of the
cell volume \textbf{(top)} and cohesive energies \textbf{(bottom)} with
several other nonlocal and dispersion-corrected density functionals.
Box plots indicate the first and third quartile values, as well as the
median (line) and mean (marked by ``$\times$''). The whiskers extend from
the box to the farthest data point within 1.5 times the inter-quartile
range from the box. Individual systems beyond this range are shown as
circles.}
\end{figure}

%%%%%%%%%%%%%%%%%%%%%%%%%%%%%%%%%%%%%%%%%%%%%%%%%%%%%%%%%%%%%%%%%%%%%%%
\subsection{X23 and G60}

In Fig.~\ref{fig:X23_boxes}, we show a detailed summary of DF3-mc's
accuracy on the X23 set and comparison to some other functionals. It is
perhaps unsurprising that DF3-mc performs well for these systems, given
that they account for nearly half of our optimization set. It is,
however, still encouraging that the functional significantly surpass the
accuracy of all other van der Waals density functional optimizations.
Out of this group, DF3-mc shows the best accuracy in cohesive energies,
with a MD of only $-6$ meV, as well as unit-cell volumes at a MARD of
$0.44$\%.  Other van der Waals functionals generally overestimate
energies of cohesion and yield mixed results for the unit-cell size.
This is especially true for DF1-optB88, which was developed by
optimizing B88 exchange for the S22 set molecular dimers
\cite{klimes_2010:chemical_accuracy}. We see similar results for the
nonlocal functional VV, though its nonlocal correlation kernel is of a
fundamentally different form than the vdW-DF functionals. Like
DF1-optB88, VV and its predecessor, VV10, were optimized with respect to
binding energies in the S22
\cite{Sabatini_2013:nonlocal_van,Vydrov_2009:nonlocal_van}.  This shared
design choice makes DF1-optB88 and VV intriguing points of comparison
for the molecular crystals because they are designed based on the same
benchmark set.  Also for that reason, however, we limit our benchmarking
of those functionals to the  X23 data set. Lastly, we find that DF3-mc
provides results similar to, but somewhat surpassing, the
dispersion-corrected PBE-D3, a popular functional for molecular solids.
PBE-D3 possesses a cohesive energy  MD of $-16$ meV and a cell volume
MARD of $0.87$\% for the X23.

\begin{figure}
\includegraphics[width=\columnwidth]{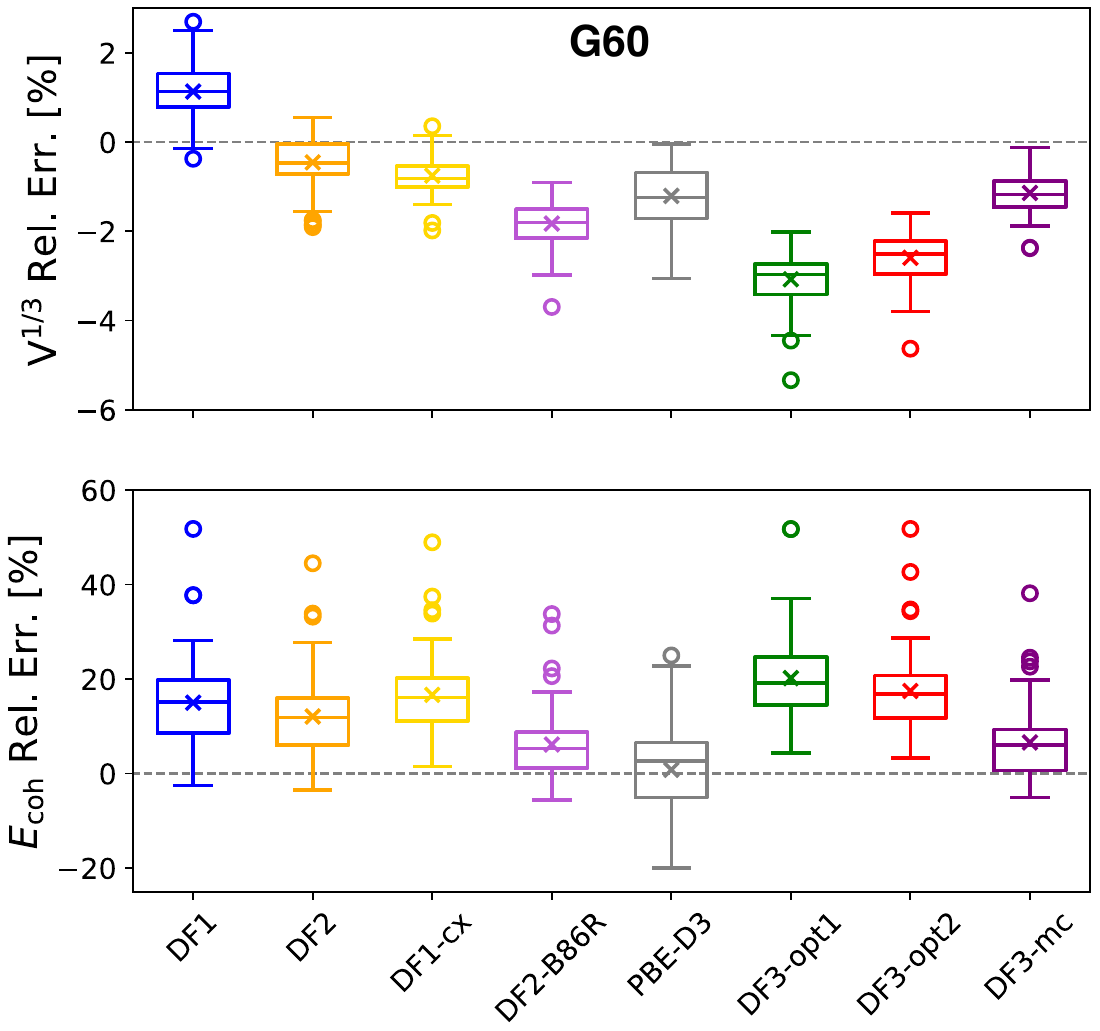}
\caption[Accuracy of DF3-mc for the G60 molecular
crystals] {\label{fig:G60_violin} Accuracy of DF3-mc on the G60 set of
molecular solids, comparing relative deviations in cohesive energies
with several other nonlocal and dispersion-corrected density
functionals. See Fig.~\ref{fig:X23_boxes} for further details.}
\end{figure}

To assess the accuracy of DF3-mc for systems outside of its optimization
set, we perform calculations on the G60 set of molecular crystals. Due
to its more diverse sampling in comparison with the X23 set, and little
overlap between the two, the G60 presents an effective way of gauging
the utility of vdW-DF-mc for broader classes of molecular crystals.
Figure~\ref{fig:G60_violin} showcases the accuracy of DF3-mc on the G60,
compared with several other vdW-DF functional and PBE-D3. We choose
these functionals primarily to compare different treatments of the
dispersion interactions, for which all four of these functionals differ.
For cohesive energies we find that the performance of DF2-B86R, PBE-D3, and DF3-mc is
superior to all other tested functionals, with a MARD of $7$--$8\%$. For the
cube root of unit-cell volumes, only DF2 and DF1-cx were more accurate than DF3-mc with its MARD of only $1.14$\%.

%%%%%%%%%%%%%%%%%%%%%%%%%%%%%%%%%%%%%%%%%%%%%%%%%%%%%%%%%%%%%%%%%%%%%%%
\subsection{POLY59, ICE10, and DMC-ICE13}

\begin{figure}[t]
\includegraphics[width=\columnwidth]{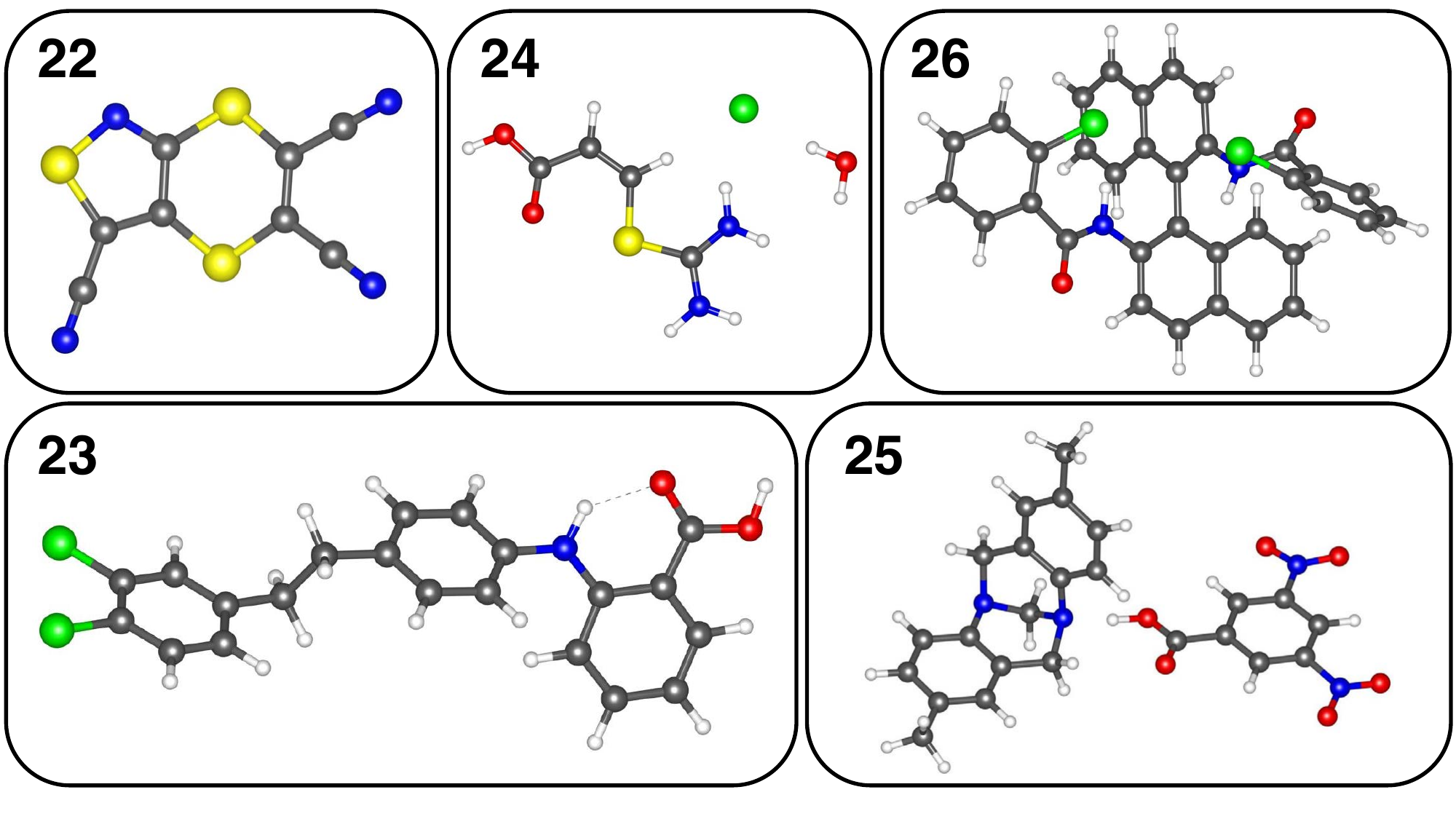}
\caption {\label{fig:POLY59_sys} Constituent molecules of each target
polymorph in the POLY59 data set. Atomic species are color-coded in
white (H), dark gray (C), blue (N), red (O), yellow (S), and green
(Cl).}
\end{figure}

An important application of DFT is the study and prediction of molecular
crystal polymorphs. The existence of multiple stable configurations of
chemically identical materials poses an immense challenge in
computational physics, and in some cases requires sub-chemical
accuracy---on the order of 1 kcal/mol---to adequately predict
\cite{nyman_2015:static_lattice}.  With the POLY59 test set of
polymorphs, we can gain some idea for how DF3-mc fares for these
scenarios. This data set is comprised of five molecular crystal species:
Tricyano-1,4-dithiino[c]-isothiazole as target 22 and
2-((4-(3,4-dichlorophenethyl)phenyl)amino)benzoic acid as target 23.
Target 24 is a chloride salt hydrate of
(Z)-3-((diaminomethyl)thio)acrylic acid. Target 25 is multi-component,
consisting of 3,5-dinitrobenzoic acid and
2,8-dimethyl-6$H$,12$H$-5,11-methanodibenzo[$b$,$f$][1,5]diazocine. And
lastly, target 26 is
$N$,$N'$-([1,$1'$-binaphthalene]-2,$2'$-diyl)bis(2-chlorobenzamide).
These five constituent molecules/complexes are shown in
Fig.~\ref{fig:POLY59_sys}.

\begin{figure}
\includegraphics[width=\columnwidth]{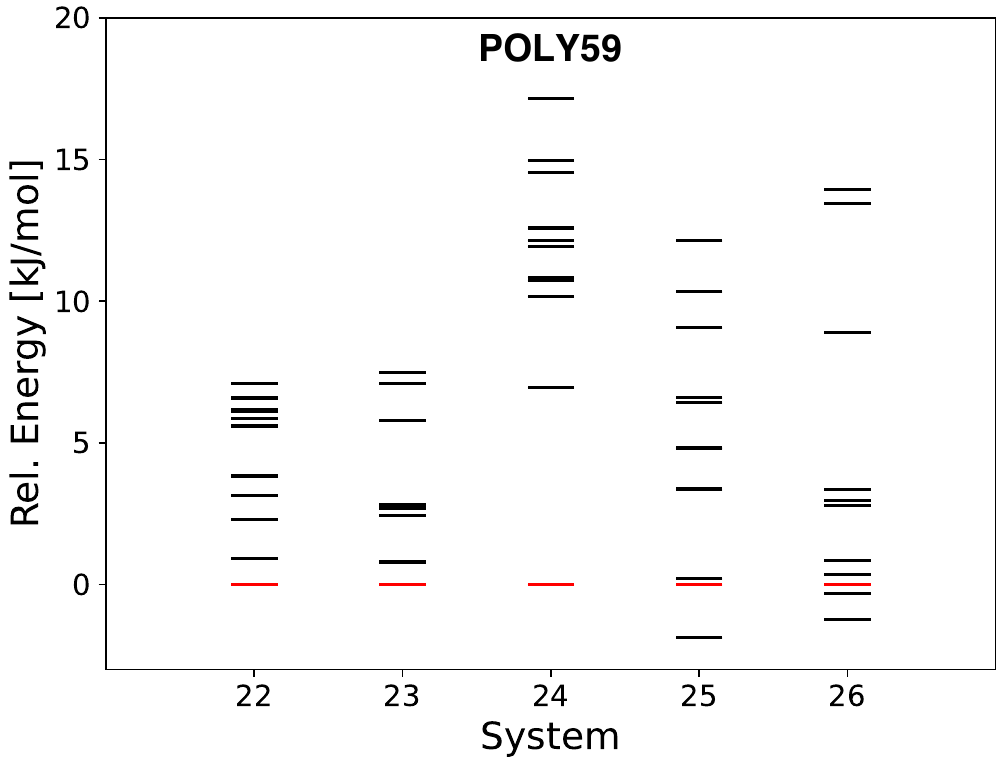}
\caption {\label{fig:POLY59} Relative energies for five different
polymorphic compounds, as calculated with DF3-mc. Red lines represent
the calculated energies of experimentally-realised polymorphs, and all
other configurational energies are taken relative to these values.}
\end{figure}

Figure~\ref{fig:POLY59} shows the computed cohesive energies for each of
the five compounds in this set, with experimental ground-state
configurations indicated by the horizontal axis. We find that DF3-mc
correctly predicts the ground state in three out of the possible five
cases: target systems 22, 23, and 24. The number of these correct
predictions, denoted as ``hits'', can be compared directly with some
other functionals that are examined in
Ref.~\cite{Brandenburg_2016:organic_crystal}. DF2, for example, achieves
only two hits: systems 22 and 24. PBE-D2 performs comparably to DF3-mc,
with hits in systems 22, 24, and 26. The Tkachenko-Scheffler (TS)
\cite{Tkatchenko_2009:accurate_molecular} method performs better still,
with hits on every system except target 26. TPSS-D3
\cite{Tao_2003:climbing_density} and PBE-D3 show the best possible
performance, each achieving hits on all five target systems. Within the
POLY59 set, target 24 possesses a large gap between the ground-state and
secondary polymorphs, explaining why so many methods successfully
predict its behavior. Target system 23 possesses five
experimentally-realized polymorphs, denoted as 23-00$\alpha$,
23-00$\beta$, and so on up to 23-00$\epsilon$. For this case, predicting
any one of the five as the ground-state would be considered a hit, and
DF3-mc specifically finds the lowest energy for 23-00$\beta$. This same
prediction was also yielded by PBE-TS, TPSS-D3, and PBE-D3, indicating
good agreement between DF3-mc and the most accurate dispersion-corrected
methods.

For the target systems that DF3-mc did not hit upon, 25 and 26, further
information can be taken from which systems were incorrectly assigned
lower energies. In target 25, system 25-02 was calculated as being 1.88
kJ/mol more favorable than the experimentally-realized polymorph, 25-00.
In fact, DF2 does the same, and with a comparable margin of error at
1.67 kJ/mol. And PBE-D3, though correctly hitting upon the ground state
of this system, still yields an energy difference of only 1.34 kJ/mol
between 25-00 and 25-02. For target 26, DF3-mc predicts 26-01 and 26-02
as being lower than the experimentally-realized structure, with a
maximum difference of 1.22 kJ/mol. This is again reflected in the
performance of DF2, which finds 26-02 to be 1.55 kJ/mol lower than
26-00.

\begin{figure*}
\includegraphics[width=0.7\textwidth]{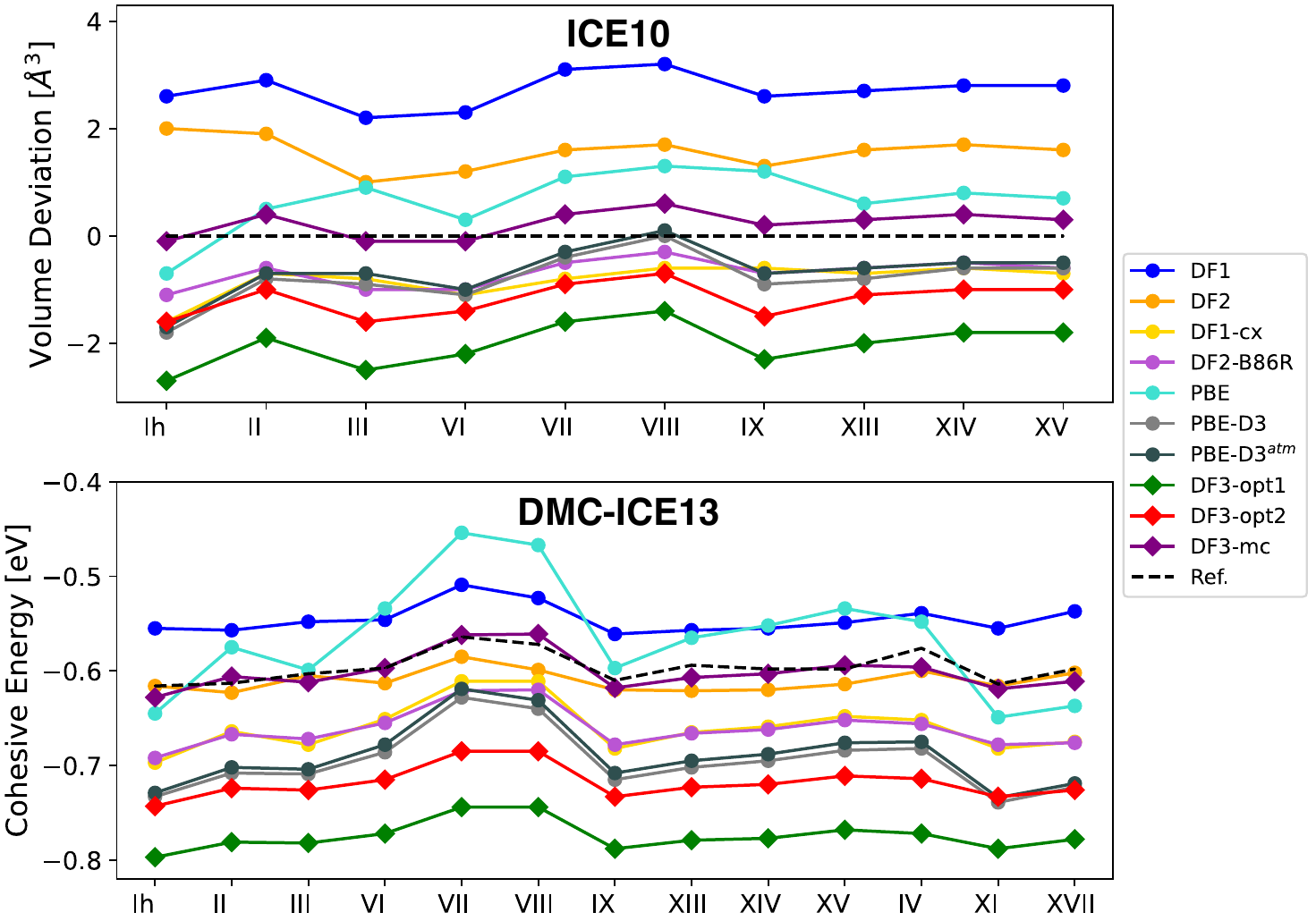}
\caption {\label{fig:df3-mc_ice} Calculated deviations in volume per
water molecule (\AA$^3$) and cohesive energies (eV) for the ICE10 and
DMC-ICE13 sets, respectively. ICE10 data for PBE, PBE-D3, and
PBE-D3$^{atm}$ is taken from taken from
Ref.~\cite{Brandenburg_2015:benchmarking_dft}, which compiles reference
volumes found via neutron diffraction experiments. DMC-ICE13 data for
PBE, PBE-D3, and PBE-D3$^{atm}$, DF1, and DF2 is taken from
Ref.~\cite{DellaPia_2022:dmc-ice13_ambient}, and contains all of the
polymorphs in the ICE10 set, plus IV, XI, and XVII to account for one of
each of the hydrogen ordered-disordered pairs. All data for individual
systems is available in the Supporting Materials.}
\end{figure*}

To continue our study of polymorphism and gauge accuracy for
H-bond-dominated systems, we examine several polymorphs of ice.  A
statistical summary of our ICE10 and DMC-ICE13 calculations is presented
in Fig.~\ref{fig:df3-mc_ice}.  We find that for these sets, DF3-mc
demonstrates excellent accuracy with respect to both the cohesive energy
and cell dimensions of ice. This is, in large part, a success of the
vdW-DF framework, as the bottom panel of Fig.~\ref{fig:df3-mc_ice} shows
that other vdW-DF's generally display very accurate trends in binding
energy, but are shifted to higher or lower overall cohesion. Minor
exceptions to this can be seen in DF1 and DF2, which have flatter
distributions than subsequent vdW-DFs. These are  also the only two
tested functionals that predict stronger binding for ice II than the
ground state, ice Ih.  This figure also demonstrates how ice may pose a
challenge for force-field dispersion corrections. For PBE-D3, with or
without the Axilrod-Teller-Muto (ATM) three-body-term
\cite{axelrod_1943:interactions_van, muto_1943:force_nonpolar}
(PBE-D3$^{atm}$), the variance in cohesive energy is noticeably larger
than any vdW-DF. In particular, the difference between ice Ih and the
high lying polymorphs VII and VIII is more exaggerated, and they both
display an overestimation of binding in ice XI and XVII that is not
present in the vdW-DFs or diffusion Monte Carlo calculations. These
characteristics are also present in uncorrected PBE, and to a greater
extent.  We find the performance of DF3-mc to be particularly promising,
because it is accurate in both trends between polymorphs and absolute
cohesive energies.  We attribute this success in modeling to our
optimization set, which include a diverse sample of hydrogen-bonded
complexes, particularly those with multiple points of interaction.

%%%%%%%%%%%%%%%%%%%%%%%%%%%%%%%%%%%%%%%%%%%%%%%%%%%%%%%%%%%%%%%%%%%%%%%
\section{Conclusions}\label{vdw-df3-mc:conclusions}
%%%%%%%%%%%%%%%%%%%%%%%%%%%%%%%%%%%%%%%%%%%%%%%%%%%%%%%%%%%%%%%%%%%%%%%

We have presented a new optimization of the third generation van der
Waals density functional, which we call vdW-DF3-mc. Using lessons
learned from prior studies of gradient contributions to exchange, we
created a novel form of the exchange enhancement factor, prioritizing
the flexibility needed to target desirable features of molecular solids'
binding profile. We also re-optimize the vdW-DF3 nonlocal correlation
with a new constraint on $q_0(\textbf{r})$, which yields back vdW-DF3's
intended flexible design. With these combined innovations, and an
extensive optimization set that includes molecular solids, dimers, and
layered structures, vdW-DF3-mc achieves extraordinarily accurate
energies and cell geometries for a variety of systems. This includes not
just solids within our test set, like the X23, but a large number of
other systems like the G60, POLY59, and ICE10 benchmark sets.  The
particular accuracy with respect to ice polymorphs, we credit to our
inclusion of hydrogen-bonded dimers within our optimization set, which
may have helped diversify the range of bond characters captured by
vdW-DF3-mc, as well as the emphasis on minimizing stress rather unit
cell deviations.  This diversification also gives vdW-DF3-mc excellent
suitability for systems with water and the ever-growing family of HOFs.
That said, we emphasize that this functional is intended for practical
modeling of molecular crystals at finite temperature under typical
experimental separations, as opposed to attempting to describe the
purely electronic problem, which would require accounting for zero-point
and thermal expansions.  Finally, we would like to note that the
framework and design strategy used here, including stress-based
functional optimization, and targeting specific electronic density
domains, can be adopted for other types of systems, through which
additional, highly-accurate nonlocal functionals may be created.

%%%%%%%%%%%%%%%%%%%%%%%%%%%%%%%%%%%%%%%%%%%%%%%%%%%%%%%%%%%%%%%%%%%%%%%%
\section*{Acknowledgement}
%%%%%%%%%%%%%%%%%%%%%%%%%%%%%%%%%%%%%%%%%%%%%%%%%%%%%%%%%%%%%%%%%%%%%%%%

TT acknowledges support from the U.S.\ National Science Foundation grant
No.\ DMR-1712425, which provided the insights necessary for the
developments presented here.  The computations in this work were done on
the high-performance cluster Saga and Fram, managed by UNINETT Sigma2.

%%%%%%%%%%%%%%%%%%%%%%%%%%%%%%%%%%%%%%%%%%%%%%%%%%%%%%%%%%%%%%%%%%%%%%%%
%bibliography{references}

%%%%%%%%%%%%%%%%%%%%%%%%%%%%%%%%%%%%%%%%%%%%%%%%%%%%%%%%%%%%%%%%%%%%%%%%
\end{document}